\documentclass[prc,aps,a4paper,groupedaddress,superscriptaddress,nofootinbib,showpacs
,preprintnumbers,twocolumn]{revtex4}
\usepackage{graphicx}
\usepackage{amsfonts}
\usepackage{amssymb,color}
\usepackage{amsmath}
\usepackage{natbib}
\usepackage{dcolumn}
\usepackage{bm} 
\newcommand{\bwt}{\begin{widetext}}
\newcommand{\ewt}{\end{widetext}}
\newcommand{\beq}{\begin{equation}}
\newcommand{\eeq}{\end{equation}}
\newcommand{\bea}{\begin{eqnarray}}
\newcommand{\eea}{\end{eqnarray}}
\begin{document}
\title{
Interplay between the symmetry energy and the strangeness content of neutron stars
}

\author{Constan\c ca Provid\^encia}
\email{cp@teor.fis.uc.pt}
\affiliation{Centro de F\'{\i}sica Computacional, Department of Physics, University of Coimbra, 3004-516 Coimbra, Portugal} 
\author{Aziz Rabhi}
\email{rabhi@teor.fis.uc.pt}
\affiliation{Centro de F\'{\i}sica Computacional, Department of Physics, University of Coimbra, 3004-516 Coimbra, Portugal} 
\affiliation{Laboratoire de Physique de la Mati\`ere Condens\'ee,
Universit\'e de Tunis El-Manar, Campus Universitaire, Le Belv\'ed\`ere-1060, Tunisia}

\date{\today}
\begin{abstract}
The effect of
the density dependence of the nucleonic equation of state and the
hyperon meson couplings on the star properties, including strangeness
content, mass and radius, are studied within a relativistic mean field formalism.
It is shown that there is still lacking  information on  the
nucleonic equation of state at supra-saturation densities and on the 
hyperon interactions in nuclear matter that will  allow a clear answer 
to the question whether the mass of the pulsar J1614-2230 
could rule out exotic degrees of freedom from the interior of compact stars.
We show that some star properties are affected in a similar way by the density 
dependence of the symmetry energy and the hyperon content of the star. To
disentangle these two effects it is essential to have a good knowledge of 
the equation of state at supra-saturation densities. A linear correlation
between the radius and the strangeness content of a star with a fixed mass is obtained.
\end{abstract}
\maketitle

\section{Introduction}

The measurement of the mass of the mili-second pulsar PSR J1614-2230
with high precision, $M=1.97\pm 0.04\, M_\odot$ \cite{demorest},
 has given origin to a large number of studies that
discuss whether there is room for hyperons or other exotic degrees of
freedom inside this star
\cite{isaac11,haensel11,schaffner12a,schaffner12b,sedrakian12,panda12}. In
particular microscopic calculations using the non-relativistic
Brueckner-Hartree-Fock formalism predict a very large softening of
the equation of state (EOS) when hyperons are taken into account, and are not even able to
account for the standard neutron star masses of the order $\sim 1.4\,
M_\odot$ \cite{bhf}. The possibility that three-body-hyperonic forces could solve
this problem was discussed in \cite{isaac11} by complementing the BHF
calculation with a density dependent Skyrme like  term based on the
model \cite{gal97} which mimics the many-body-term. Reasonable
assumptions seem to exclude the appearance of stars with masses larger
than $1.6\, M_\odot$. It would be important to verify whether within a
relativistic microscopic approach the same conclusions would be
drawn. It is known that among the important effects
included in the relativistic models, the  saturation of the scalar
field gives rise to
effects that can be interpreted as many-body effects and could give
rise to the extra repulsion needed \cite{walecka97}.

Hyperons may appear in the
inner core of neutron stars at densities of about 2 $-$ 3$\rho_0$ and
have first been introduced in relativistic mean-field nuclear models 
in \cite{hyp1,hyp2,hyp3}. In particular, in reference \cite{hyp3}
the strange mesons $\sigma^*$ and $\phi$  have also been introduced. We will fix the couplings of the hyperons
to mesons following a similar  way to the one proposed in~\cite{hyp3}.

 In \cite{schaffner12a} the authors
have shown 
that it was essential to include the
strange meson $\phi$ in order to get
 stiff enough EOS that could describe a star with a mass of $\sim
2M_\odot$. The calculation was done within a relativistic mean-field
(RMF) approach  including only  non-linear $\sigma$-meson terms. The
authors have concluded that the compression modulus  at saturation had
little effect on the maximum mass, which,  however, was very sensitive
to the effective mass. Similar conclusions with respect to the EOS
had been  drawn in
\cite{greiner88}. Within this approach a small effective mass at
saturation requires a large sigma meson coupling  that has to be
compensated by a large nucleon-omega meson coupling,
giving rise to large vector contributions at high densities, and,
therefore, a stiff EOS and large star masses. If, however, constraints
on the EOS like the ones coming from the flow analysis of nuclear
matter in heavy ion collisions \cite{danielewicz}  are imposed,
stiff EOS at high densities are ruled out. The softening of the EOS at
large densities may be achieved including a quartic term on the
$\omega$-meson as in \cite{tm1} or density dependent couplings as in
\cite{tw}. 

In the present study we will analyze how the content of
strangeness of a star is defined by  the properties of the nucleonic
EOS and the hyperon-meson couplings.  In particular
we will discuss: a) the effect of the density dependence of the
symmetry energy, b) the effect of the incompressibility of the EOS at
2-3 nuclear saturation density, c) the joint effect of these two
properties together with the uncertainty on the hyperon couplings.
The effect of the strange content on the maximum mass will be discussed.
It will be shown not only that the presence of the strange mesons
$\sigma^*$ and $\phi$ is important to give rise to a hard hyperonic EOS
at large densities as already shown in
\cite{schaffner12a,haensel11}, but also that the density dependence of
the EOS at 2-3 times saturation density has noticeable effects on the
star properties. In \cite{sagert2012}, the authors study the implication of a soft EOS, 
as obtained from kaon production in heavy-ion collisions at these same densities \cite{kaos}, 
on the properties of compact star.

In order to be able to quantize the effects we take as reference the TM1
parametrization of relativistic mean-field  nuclear models
\cite{tm1}. This model was fitted to the ground-state properties of
several  nuclei and to the Dirac-Brueckner-Hartree-Fock calculations
at large densities. This last feature  was only possible with the insertion of a
quartic self-interaction $\omega$ term which softens the vector meson
field at high densities. The effect of this softening on the hyperon
content will be discussed by varying the strength of the quartic
term. The slope of the symmetry energy of TM1 at saturation $L=110$
MeV is too
high according to the present experimental constraints coming from
different nuclear properties, lying close to the upper limit of
isospin diffusion in heavy ion collisions \cite{chen2005}. In order to
test the effect of the symmetry energy slope we will introduce a
non-linear $\omega-\rho$ term as in \cite{hor01} and change the
coupling so that  models with equivalent isoscalar
properties of TM1 but different density dependence of the symmetry
energy will be obtained. As discussed in
\cite{hor01,hor03,rafael11,panda12} this term reflects itself on
the star properties, namely giving rise to a smaller radii. We
will show that this effect is larger for nucleonic stars than for hyperonic
stars and that, in fact,  the presence of hyperons gives rise to a
similar effect that masks the symmetry energy one.

In section \ref{form} we present the formalism used, in section
\ref{results} the results are presented and discussed and finally in
section \ref{conclusion} some conclusions are drawn.

\section{Formalism} \label{form}
To describe the hadronic matter, we employ a relativistic mean-field (RMF) 
approach, in which the baryons interact via the exchange of mesons. The baryons considered in this work are nucleons ($n$ and $p$) and hyperons ($\Lambda$, $\Sigma$, and $\Xi$). The exchanged mesons include scalar and vector mesons ($\sigma$ and $\omega$), isovector meson ($\rho$), and two additional hidden-strangeness mesons ($\sigma^*$ and $\phi$). The Lagrangian density includes several non-linear terms in order to describe adequately the saturation properties of nuclear matter. For neutron star matter consisting of neutral mixture of baryons and leptons in $\beta$-equilibrium, we start from the effective Lagrangian density of the non-linear Walecka model (NLWM) \cite{tm1}  
\bea
{\cal L}_{NLWM}&=&\sum_{B}\bar{\Psi}_{B} \left[\gamma_{\mu}D^{\mu}_{B}- m^{*}_{B}\right]\Psi_{B} \cr
&+& \sum_{l} \bar{\psi}_{l}\left[i\gamma_{\mu}\partial^{\mu}-m_{l}\right]\psi_{l} \cr
&+&\frac{1}{2}\left(\partial_{\mu}\sigma \partial^{\mu}\sigma-m^{2}_{\sigma}\sigma^{2}\right)
-\frac{1}{3!}k\sigma^3-\frac{1}{4!}\lambda\sigma^4 \cr
&+&\frac{1}{2} m^{2}_{\omega}\omega_{\mu}\omega^{\mu}
-\frac{1}{4} \Omega_{\mu \nu} \Omega^{\mu \nu}
+\frac{1}{4!}\xi g_{\omega}^4 \left(\omega_{\mu}\omega^{\mu}\right)^2 \cr
&+&\frac{1}{2}m^{2}_{\rho}\boldsymbol{b}_{\mu}\boldsymbol{b}^{\mu}
-\frac{1}{4} \mathbf{B}_{\mu \nu} \mathbf{B}^{\mu \nu}\cr
&+&\Lambda_{\omega}\left(g^{2}_{\omega} \omega_{\mu}\omega^{\mu}\right)\left(g^{2}_{\rho} \boldsymbol{b}_{\mu}.\boldsymbol{b}^{\mu}\right)  \cr
&+&\frac{1}{2}\left(\partial_{\mu}\sigma^* \partial^{\mu}\sigma^*
-m^{2}_{\sigma^*}{\sigma^*}^{2}\right) \cr
&+&\frac{1}{2} m^{2}_{\phi}\phi_{\mu}\phi^{\mu}
-\frac{1}{4} \phi_{\mu \nu}\phi^{\mu \nu}
\label{lagran}
\eea
where $D^{\mu}_{B}=i\partial^{\mu}-g_{\omega  B} \omega^{\mu}-g_{\phi  B} \phi^{\mu}-g_{\rho B}\boldsymbol{\tau}_{B}.\boldsymbol{b}^{\mu}$ and $m^{*}_{B}=m_{B}-g_{\sigma B}\sigma-g_{\sigma^* B}\sigma^*$ is the baryon effective mass. $\Psi_{B}$ and $\psi_{l}$ are the baryon and lepton Dirac fields, respectively, and $\sigma$, $\omega$, and $\rho$ represent the scalar, vector, 
and vector-isovector meson fields, which describe the nuclear interaction. 
The coupling constants of mesons $i=\sigma, \omega, \rho$ with baryon $B$ are denoted by 
$g_{i,B}$ where the index $B$ runs over the eight lightest baryons $n$, $p$, 
$\Lambda$, $\Sigma^-$, $\Sigma^0$, $\Sigma^+$, $\Xi^-$ and $\Xi^0$,
and the sum on $l$ is over electrons and muons ($e^{-}$ and $\mu^{-}$).  
The baryon mass and the lepton mass are denoted by $m_{B}$ and
$m_{l}$, respectively. The constants $k$ and $\lambda$ are the weights of the scalar self-interaction terms and $\boldsymbol{\tau}_{B}$ is the isospin operator. The mesonic field tensors are given by their usual expressions: 
$\Omega_{\mu \nu}=\partial_{\mu}\omega_{\nu}-\partial_{\nu}\omega_{\mu}$,
$\boldsymbol{B}_{\mu \nu}=\partial_{\mu}\boldsymbol{b}_{\nu}-
\partial_{\nu}\boldsymbol{b}_{\mu}$, and $\phi_{\mu \nu}=\partial_{\mu}\phi_{\nu}-\partial_{\nu}\phi_{\mu}$.

In the RMF model, the meson fields are treated as classical fields,
and the field operators are replaced by their expectation
values. Applying the Euler-Lagrange equations to Eq.(\ref{lagran}) and
using the mean-field approximation, we obtain the following meson field
equations of motion as following, with $g_{iB}=x_{iB} g_{i}$,
\begin{eqnarray}
\sigma_0&=&
\frac{g_\sigma}{m^{2}_{\sigma,eff}} \sum_{B}\frac{x_{\sigma
    B}}{\pi^2}\int_{0}^{k^{B}_{F}}\frac{m^*_B\, k^2\, dk}{\sqrt{k^2+  {m^*_B}^2}}\\
\omega_0 &=&  \frac{g_{\omega}}{m_{\omega, eff}^2}\sum_{B} \frac{x_{\omega B}\left(k^B_
    F\right)^3}{3\pi^2} ,  \\
b_0 &=&  \frac{g_{\rho}}{m_{\rho,eff}^2}\sum_{B} \frac{x_{\rho B}{\tau}_{3 B}\left(k^B_
    F\right)^3}{3\pi^2} ,\\
\sigma^*_0 &=&  \frac{g_{\sigma^*}}{m^2_{\sigma^{*}} }\sum_{B}\frac{x_{\sigma^*
    B}}{\pi^2}\int_{0}^{k^{B}_{F}}\frac{m^*_B \,k^2 dk }{\sqrt{k^2+{m^*_B}^2} } \\
\phi_0 &= &\frac{g_\phi}{m^{2}_{\phi}} \sum_{B} \frac{x_{\phi B}\left(k^B_
  F\right)^3}{3\pi^2},
\label{mesons}
\end{eqnarray}
where
\begin{eqnarray} 
m_{\sigma,eff}^2&=& m_{\sigma}^2 +\frac{k}{2}\sigma_0+\frac{\lambda}{6}\sigma_0^2\\
m_{\omega, eff}^2&=&m_{\omega}^2+\frac{\xi}{6} g_\omega^4\omega_0^2
+2\Lambda_\omega g_\omega^2 g_\rho^2 b_0^2,\\
m_{\rho, eff}^2&=&m_{\rho}^2
+2\Lambda_\omega g_\omega^2 g_\rho^2 \omega_0^2.
\label{masseff}
\end{eqnarray}
We see that the non-linear $\omega$ and $\rho$ terms give rise to 
effective masses for the $\omega$ and $\rho$ mesons that increase with
density, giving rise to a softening of the vector fields at large densities.
In this work, we employ the TM1 parameter set of the RMF model. The meson-hyperon and the strange meson-hyperon coupling constants $g_{\omega H}$, $g_{\rho H}$, $g_{\sigma^* H}$, and $g_{\phi H} $ are determined by using SU(6) symmetry 
\bea
\frac{1}{3}g_{\omega N}&=&\frac{1}{2}g_{\omega \Lambda}= \frac{1}{2}g_{\omega \Sigma}=g_{\omega \Xi}, \cr
g_{\rho N}&=&g_{\rho \Lambda}= \frac{1}{2}g_{\rho \Sigma}=g_{\rho \Xi} \cr
2 g_{\phi \Lambda}&=& 2 g_ {\phi \Sigma}= g_ {\phi \Xi}=-\frac{2\sqrt{2}}{3}g_{\omega N} 
\eea
where $N$ means nucleon ($g_{iN} \equiv g_i$). The scalar coupling
constants are chosen to give reasonable potentials. The coupling
constants $g_{\sigma H}$ of the hyperons with the scalar meson
$\sigma$ are adjusted to the potential depths $U_H^{(N)}$ felt by a
hyperon H in symmetric nuclear matter  at saturation following the relation
\beq
U^{N}_H=x_{\omega H} V_{\omega}-x_{\sigma H} V_{\sigma} 
\eeq
with $x_{i,H}=g_{i, H}/g_{i}$, $ V_{\omega} \equiv g_{\omega}\omega_0 $ and $V_{\sigma} \equiv g_{\sigma}\sigma_0 $ are the nuclear potentials for symmetric nuclear matter at saturation density.
For the present work we will fix $U^{N}_\Lambda=-28$ MeV, and use
$U^{N}_\Sigma=-30,\; 0,\;30$ MeV, and for $U^{N}_\Xi$ we will use
different values $-18$, 0, and $18$ MeV. 
For the $\sigma^*$ meson we consider a weak YY coupling and take $U_\Lambda^\Lambda
\sim -5$ MeV. together with $2 g_{\sigma^* \Lambda} = 2 g_ {\sigma^* \Sigma}= g_ {\sigma^* \Xi}$.
All hyperon coupling ratios $\left\lbrace g_{\sigma H}, g_{\omega H}, g_{\rho H} \right\rbrace_{H = \Lambda, \Sigma, \Xi} $ are determined once the coupling constants  $\left\lbrace g_{\sigma}, g_{\omega}, g_{\rho} \right\rbrace$  of the nucleon sector are given. The hyperons masses are taken to be $m_\Lambda=1116$ MeV, $m_{\Sigma^+}=1189$ MeV, $m_{\Sigma^0}=1193$ MeV, $m_{\Sigma^-}=1197$ MeV and $m_{\Xi^0}=1315$ MeV, $m_{\Xi^-}=1321$ MeV, while the strange meson masses are $m_{\sigma^*}=980$ MeV and $ m_\phi=1020$ MeV.

\begin{table}
\caption{Coupling constants and masses for the TM1 and TM1-2 models which have the same saturation properties: saturation density $\rho_0=0.145$ fm$^{-3}$, binding energy $E/A= -16.30$ MeV, symmetry energy $J=36.93$ MeV, incompressibility $K= 281.28$ MeV, effective mass $M^*/M= 0.63$.}
\label{table2}
\begin{ruledtabular}
\begin{tabular}{ c c c c c c c c c c c c}
Model & $\left(\frac{g_{\sigma}}{m_{\sigma}}\right)^2$ & $\left(\frac{g_{\omega}}{m_{\omega}}\right)^2$ & $\left(\frac{g_{\rho}}{m_{\rho}}\right)^2$  &$k/M$ & $\lambda$ & $\xi$\\
 & $(\hbox{fm})^2$ & $(\hbox{fm})^2$ & $(\hbox{fm})^2$ &  & & \\
\hline
TM1    &  15.0125 & 10.1187 & 5.6434  & 3.0655 & 2.7333 & 0.06  \\
TM1-2 & 14.9065 & 9.9356 & 5.6434   & 3.5351 & -47.8812 & 0.04  \\
\end{tabular}
\end{ruledtabular}
\end{table}

For neutron star matter consisting of a neutral mixture of baryons and leptons, the $\beta$ equilibrium condition without neutrino trapping are given by
\bea
\mu_{p}&=&\mu_{\Sigma^+}=\mu_{n}-\mu_{e}\cr
\mu_{\Lambda}&=&\mu_{\Sigma^0}=\mu_{\Xi^0}=\mu_{n}\cr
\mu_{\Sigma^-}&=&\mu_{\Xi^-}=\mu_{n}+\mu_{e}\cr
\mu_{\mu}&=&\mu_{e}
\eea
where $\mu_i$ is the chemical potential of species $i$. The chemical potentials of baryons and leptons are given by
\bea
\mu_{B}&=&\sqrt{{k^B_F}^2+{m^*_B}^2}+g_{\omega
  B} \omega_0+g_{\rho B}\tau_{3 B} b_0  +g_{\phi B}\phi_0\nonumber\\
\mu_{l}&=&\sqrt{{k^l_F}^2+m^2_l}.
\eea
and the charge neutrality condition is written by
\beq
\rho_p+\rho_{\Sigma^+}=\rho_e+\rho_{\mu}+\rho_{\Sigma^-}+\rho_{\Xi^-},
\eeq
where $\rho_i=\left(k^{i}_{F}\right)^3/\left(3\pi^2\right)$ is the number density of species $i$. We solve the coupled equations self-consistently at a given baryon density $\rho_B=\rho_{n}+\rho_p+\rho_{\Lambda}+\rho_{\Sigma^+}+\rho_{\Sigma^0}+\rho_{\Sigma^-}+\rho_{\Xi^0}+\rho_{\Xi^-}$ 
Through the energy-momentum tensor, we obtain the total energy density and the pressure of the neutron star matter
\bea
\varepsilon&=&\sum_{B}\frac{1}{\pi^2}\int_{0}^{k^{B}_{F}}\sqrt{k^2+{m^*_B}^2}k^2
dk\cr 
&+&\sum_{l=e,\mu}\frac{1}{\pi^2}\int_{0}^{k^{l}_{F}}\sqrt{k^2+m^{2}_{l}}k^2 dk \cr
&+& \frac{1}{2}m^{2}_{\sigma}\sigma^{2}_{0}+\frac{k}{6}\sigma^3_{0}+\frac{\lambda}{24}\sigma^4_{0}
+\frac{1}{2}m^{2}_{\omega}\omega^{2}_{0}+\frac{1}{{8}} \xi
g_{\omega}^4\omega^{4}_{0}\cr
&+&\frac{1}{2}m^{2}_{\rho}b^{2}_{0}+3\Lambda_{\omega} g^{2}_{\omega} g^{2}_{\rho}\omega^{2}_{0}b^{2}_{0}\cr
&+&\frac{1}{2}m^{2}_{\sigma^*}{\sigma^*}^{2}+\frac{1}{2}m^{2}_{\phi}\phi^{2}_{0},
\eea
\bea
P&=&\sum_{B}\frac{1}{3\pi^2}\int_{0}^{k^{B}_{F}}\sqrt{k^2+{m^*_B}^2}k^2
dk\cr
&+&\sum_{l=e,\mu}\frac{1}{3\pi^2}\int_{0}^{k^{l}_{F}}\sqrt{k^2+m^{2}_{l}}k^2 dk \cr
&-& \frac{1}{2}m^{2}_{\sigma}\sigma^{2}_{0}-\frac{k}{6}\sigma^3_{0}-\frac{\lambda}{24}\sigma^4_{0}
+\frac{1}{2}m^{2}_{\omega}\omega^{2}_{0}+\frac{1}{{24}}\xi
g_{\omega}^4\omega^{4}_{0}\cr
&+&\frac{1}{2}m^{2}_{\rho}b^{2}_{0}+\Lambda_{\omega} g^{2}_{\omega} g^{2}_{\rho}\omega^{2}_{0}b^{2}_{0}\cr
&-&\frac{1}{2}m^{2}_{\sigma^*}{\sigma^*}^{2}+\frac{1}{2}m^{2}_{\phi}\phi^{2}_{0}.
\eea

\section{Results and discussions}\label{results}

\begin{figure}[hb]
\centering
\includegraphics[width=0.95\linewidth,angle=0]{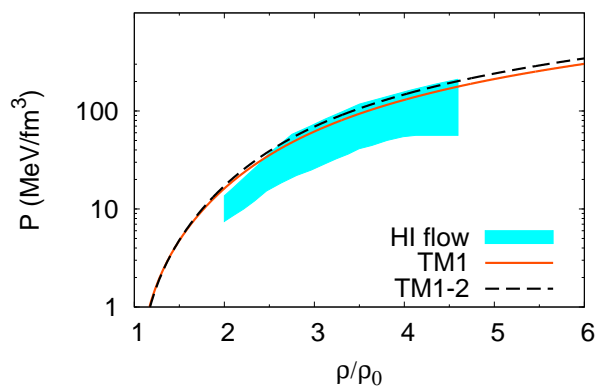}
\caption{(Color online) EOS for TM1, TM1-2 and FSU \cite{fsu} above saturation density. The shaded region
  represents constraints imposed by heavy ion flow \cite{danielewicz}.}
\label{tm1}
\end{figure}

In order to better understand the effect of the symmetry energy and
the incompressibility of the EOS at high densities together with the
uncertainty on the hyperon interaction on the strangeness content, the mass and radius of the stars,
we consider the parametrization TM1,  a parametrization that satisfies the heavy-ion flow
constraints for symmetric matter at 2-3$\rho_0$ \cite{danielewicz},
see Fig. \ref{tm1}. { Even though in the following we will use parametrizations which
satisfy this constraint, it is important to point out that,  since it is hard to model flow in transport
simulations, this constraint should be taken with care.}

We will also consider two variations of TM1:
a) we will consider the  EOS of TM1 with a smaller quartic omega
term, but with the same properties at saturation. We designate this
parametrization as TM1-2, and, as shown in Fig.~\ref{tm1}, it is stiffer than with
TM1 at supra-saturation densities but still within the constraints imposed by
heavy ion flow~\cite{danielewicz};
b) the symmetry energy slope of TM1 at saturation density is $L=110$ MeV, a value
which is presently considered too high (see~\cite{chen2005,slope}), and, therefore, we 
also take a second parametrization changing the isovector channel of TM1 by including 
a non-linear $\omega\rho$ term that makes the symmetry energy slope softer at 
supra-saturation densities. We will choose  $L=55,\, 70$ and  80 MeV. Using these  
nucleonic EoS we will test different hyperon interactions in nuclear matter.

 For the hyperon potentials in symmetric nuclear
matter at saturation we take $U^{N}_\Lambda=-28$ MeV,
$U^{N}_\Sigma=30$ MeV and $U^{N}_\Xi=\pm18$ MeV. The two values of $U^{N}_\Xi$
take into account some  uncertainty on the experimental data on this potential
\cite{gal2010}. Finally, we 
also consider the inclusion of the strange  mesons
$\sigma^*,\,\phi$. 
According to recent experimental    
$\Lambda-\Lambda$-hypernuclear data, the $\Lambda-\Lambda$ interaction is
 only weakly attractive \cite {gal2011}. 
The effect of the small attractiveness of the 
hyperon-hyperon coupling will be considered by choosing a) a weak
$g_{\sigma^* H}$ coupling; b)   the extreme value $g_{\sigma^* H}=0$.

\begin{figure}[ht]
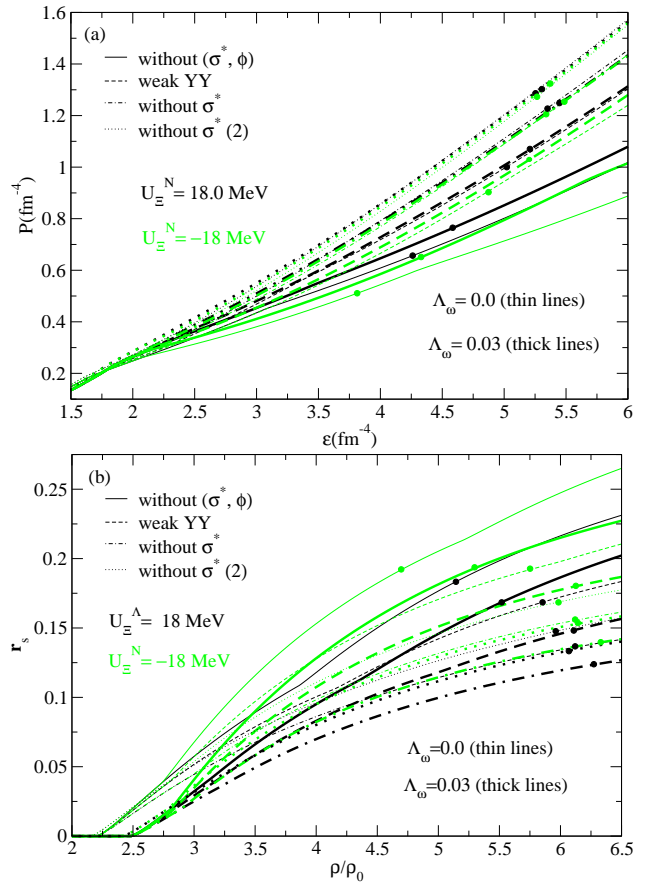

\centering
\includegraphics[width=0.95\linewidth,angle=0]{fig2a.eps}\\
\vspace{0.0cm}
\includegraphics[width=0.95\linewidth,angle=0]{fig2b.eps}
\caption{(Color online) Pressure versus energy density (top) and strangeness fraction versus
  baryonic density (bottom) for TM1 model, for different
  values of $U^{N}_{\Xi}$ (-18 MeV green curves and +18 MeV black
  curves) and the $\Lambda_{\omega}$ coupling (0 corresponding to
  $L=110$ MeV thin curves and 0.03 for $L=55$ MeV thick lines). The
  dots identify the maximum mass configuration. The dotted curves
  correspond to smaller quartic omega term.}
\label{eos}
\end{figure}
In Fig.~\ref{eos}, top panel, the EOS for the different choices of the nucleonic and
hyperonic parametrizations are shown. The full dots indicate the central density
of the most massive stable stars within each parametrization. 
We do not include the EOS without hyperons in the figure in order not to
burden it, but, as shown in Fig.~7 of~\cite{providencia12} it corresponds to
the hardest EOS and gives rise to a maximum star mass of 2.18M$_\odot$ for $L=110$ MeV
and 2.13$M_\odot$ if $L=55$ MeV (see Table \ref{tab1}).
The full lines describe the EOS
without the strange mesons and these are the softer EOS. In this case the
difference between the $L=55$ MeV (thick lines) and $L=110$ MeV (thin lines)
is the largest. This is also the situation when the $\Xi$ potential has the largest
effect. The strangeness content corresponding to these EOS, see  bottom panel
of Fig.~\ref{eos}, explains the differences: the onset of strangeness occurs
at lower densities for $L=110$ MeV and increases faster. A more attractive
$\Xi$ potential also gives rise to larger strangeness fractions.  An
interesting effect is that although having a softer EOS, stars whose
strangeness fraction increase faster with density have lower central
densities, as if the star did not support more than a given strangeness
content. 

As already discussed in \cite{rafael11,panda12}, a smaller slope
$L$ implies a softer increase of the strangeness fraction with density.
However, because the central density of these stars is larger, it is important
to study the total hyperon content of the star. This will be done by calculating for each star the total
strangeness number
$$N_S=4\pi\int_0^R \frac{\rho_s\, r^2}{\sqrt{1-2m(r)/r}} dr,$$
where $m(r)$ is the mass inside the radius $r$.

Including strange mesons
washes out the effect of the symmetry energy and of the $\Xi$ optical potential on the EOS, 
mainly if only the $\phi$ meson, that gives rise to extra  repulsion between hyperons, is
included (dash-dotted lines).

Taking the nucleonic TM1-2 EOS,  a stiffer EOS than TM1 at large densities
(dotted lines) the EOS remains always stiffer than TM1
 although having a larger strangeness content.

\begin{figure}[ht]
\centering
\includegraphics[width=1.\linewidth,angle=0]{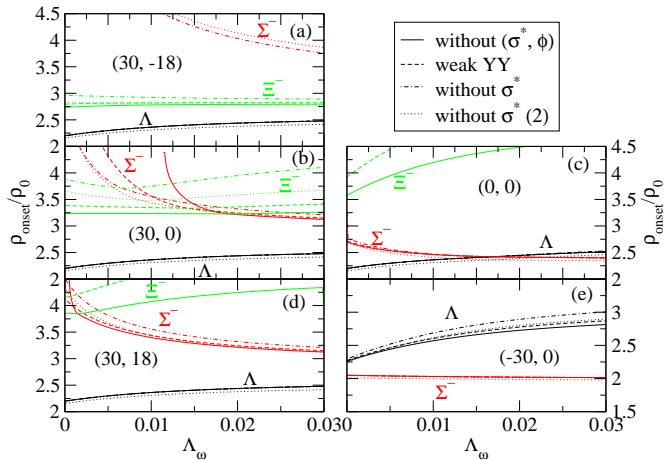}
\caption{(Color online) Onset of the $\Lambda$, the $\Sigma^-$, and the $\Xi^-$
  hyperons as a function of the parameter $\Lambda_{\omega}$ and with
  $U^N_\Lambda=-28$ and several values of $U^N_\Sigma$ and $U^N_\Xi$
  in symmetric nuclear matter at saturation. The limiting values of
  $\Lambda_{\omega}$ correspond to 110 MeV (0) and 55 MeV (0.03). The
  pair of values given in each graph refers to $(U^N_\Sigma,U^N_\Xi)$.}
\label{onset}
\end{figure}

We have seen that the symmetry energy affects the onset of hyperons,
namely postponing to larger densities the onset of hyperons for smaller $L$
values. In Fig.~\ref{onset} it is shown that the different hyperons are
affected in a different way by the symmetry energy. In this figure we plot as
a function of the coupling $\Lambda_\omega$, $\Lambda_\omega=0$ (0.03) corresponds
to 110 (55) MeV, the onset of the $\Lambda$, $\Sigma^-$ and $\Xi^-$. It is
seen that the onset of $\Sigma^-$ always decreases with the decrease of $L$,
due to its larger isospin. On the other hand, the onset of $\Lambda$ occurs at
larger densities. The $\Xi^-$ never is the first hyperon to appear due to its
large mass, but, according to the attractiveness of its potential in nuclear
matter, it can appear as the second hyperon. If the repulsiveness of the
$\Sigma^-$ in nuclear matter is confirmed  we may expect that the $\Lambda$ is
the first hyperon to set on and, therefore, with a smaller slope $L$ the onset
of strangeness occurs at larger densities. However, if the optical potential
of the $\Sigma^-$ in nuclear matter is only slightly repulsive there may be a
competition between the onset of $\Lambda$ and $\Sigma^-$ depending on the
$L$, with smaller values of $L$ favoring the $\Sigma^-$ hyperon (see top
figure of the right column).

\begin{figure}[ht]
\vspace{0cm}
\centering
\includegraphics[width=0.95\linewidth,angle=0]{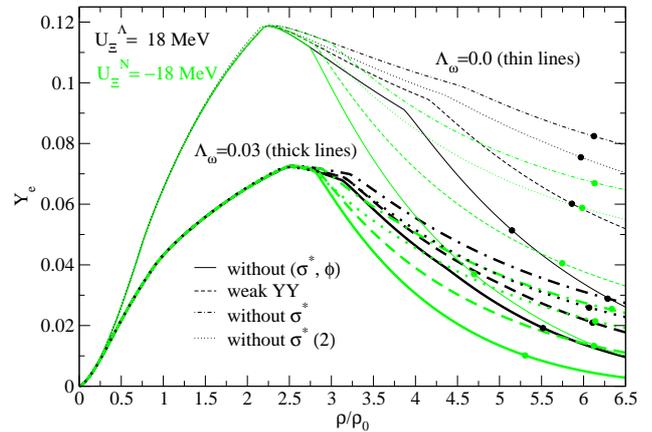}
\caption{(Color online) Electrons fraction $Y_e=\rho_e/\rho_0$ as function of the
  ratio $\rho/\rho_0$ for TM1 model and different values of
  $U^{N}_{\Xi}$ and the $\Lambda_{\omega}$ coupling (see caption of
  Fig. \ref{eos}). The dots identify the maximum mass configuration.}
\label{elefrac}
\end{figure}

Lepton fractions are also strongly affected by the symmetry energy density
dependence and the hyperon interaction. From the equations of motion for the
mesons, Eqs.~(\ref{mesons}) and (\ref{masseff}), it is seen that the non-linear
$\omega-\rho$ term gives rise to an effective mass for the $\rho$-meson that
increases with density, giving rise to a weaker $\rho$-meson field,
Eq. (\ref{mesons}). An immediate consequence is  a smaller 
asymmetry proton-neutron term in the total energy and larger allowed
differences between neutrons and protons. Smaller proton fractions
give rise to smaller electron fractions for smaller slopes $L$, as is clearly
seen in Fig.~\ref{elefrac}. Including hyperons will further reduce the electron
fraction mainly if the hyperon couplings favor the appearance of negatively
charged hyperons. This explains the difference between the green curves with
$U^{N}_\Xi=-18$ MeV which favors the onset of $\Xi^-$, with respect to the black
curves. The onset of neutrally charged hyperons also reduces the electron
fraction although not so strongly. For $L=110$ MeV the electron fraction at
the center of the star can go from 0.04 to 0.1 depending on the hyperon
interaction. This uncertainty reduces to a fraction between 0.01 and 0.03 for
$L=55$ MeV. Smaller electron fractions are generally connected to larger
neutrino fractions in matter with trapped neutrinos and therefore, the cooling
in the early seconds of a proton-neutron star 
will be strongly affected by the slope $L$.

\begin{figure}[ht]
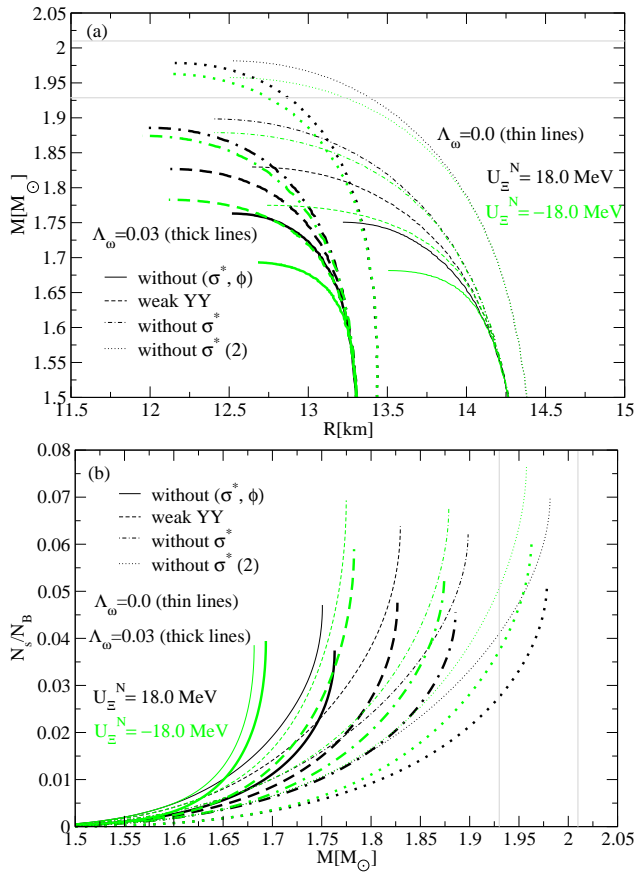

\vspace{0cm}
\centering
\includegraphics[width=0.95\linewidth,angle=0]{fig5a.eps}\\
\vspace{0cm}
\includegraphics[width=0.95\linewidth,angle=0]{fig5b.eps}
\caption{(Color online) Gravitational mass versus the radius (top panel) and  strangeness
number over the total baryonic number versus gravitational mass (bottom
panel) for TM1, and different values of $U^{N}_{\Xi}$ and the
$\Lambda_{\omega}$ coupling. The lines indicate the lower and upper limit of the PSR
J1614-2230 mass.}
\label{MR}
\end{figure}

 By solving the Tolman-Oppenheimer-Volkoff equations~\cite{tov},
  resulting from Einstein's general relativity equations for
  spherically symmetric and static stars, the neutron star profiles
  are obtained from the EOS studied, for severals values of
  $U^{N}_{\Xi}$ and the $\Lambda_{\omega}$ coupling. { For
    the outer crust EOS and the bottom inner crust EOS we
    consider the BPS EOS \cite{bps}.}

We now analyze the gravitational mass/radius curves of the families of stars 
described by the above EOS and their strangeness content (Fig.~\ref{MR} and
Table~\ref{tab1}). In the bottom panel of the
Fig.~\ref{MR} we plot, as a function of the gravitational mass, the total
strangeness number of the star over the total baryonic  number, which measures
the total strangeness content of the star. The strangeness degree of freedom
is only present in stars with gravitational masses above 1.5 $M_\odot$,
{ and the strangeness content generally attains larger values for $L=110$ MeV (thin lines).}

\begin{figure}[ht]
\centering
\includegraphics[width=1.0\linewidth,angle=0]{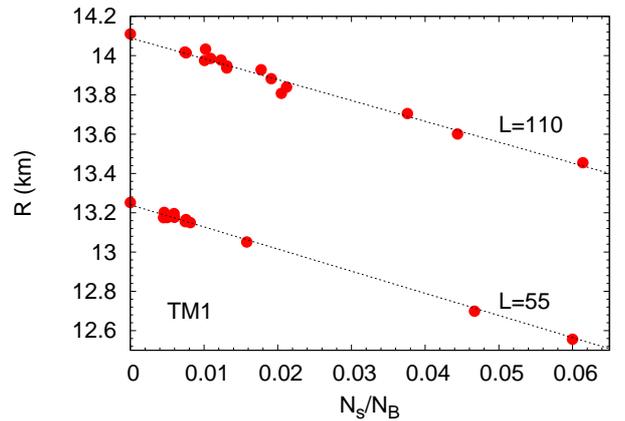}
\caption{(Color online) Radius of a star with mass $1.67\, M_\odot$ as a function of the
  strangeness content, for TM1 with $L=110$ MeV (top line with slope -11.27
  $\pm 4\%$ km) and TM1 with
  $\omega\rho$ term and $L=55$ MeV (bottom line with slope -10.62 $\pm 1\%$ km).}
\label{svr}
\end{figure}

In order to help the analysis of this information,
in
  Fig. \ref{svr} we plot the radius of a star with a mass 1.67$M_\odot$
  similar to the mass of 
the pulsar PSR J1903+0327 ($1.67\pm 0.02 M_\odot$) \cite{j1903} as a function of
its strangeness content. The largest strangeness fractions were obtained
considering an attractive potential for the $\Sigma^-$ meson. It is
interesting to notice that two almost parallel straight lines are obtained:
for $L=110$ MeV the  slope  is -11.27  $\pm 4\%$ km and
  for $L=55$ MeV  the slope is  -10.62 $\pm 1\%$ km. The straight
lines cross the vertical axis for a nucleonic star with no hyperons.
The slope is almost independent of $L$.

\begin{figure*}[ht]
\centering
\includegraphics[width=0.95\linewidth,angle=0]{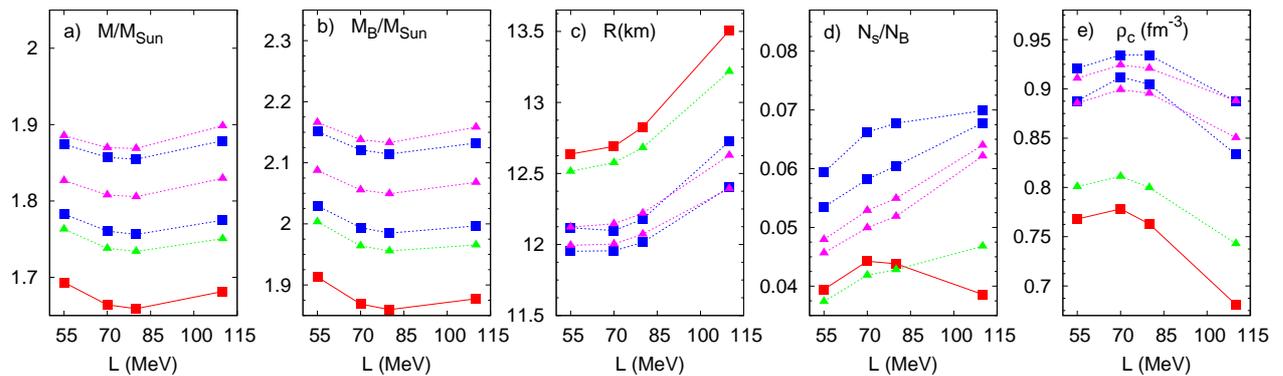}
\caption{(Color online) 
Properties of maximum mass stars obtained with $L=55, 70, 80$ and 110 MeV,
for TM1:  $U^{N}_\Xi=+18$ MeV (triangles)  without (green) and  with (pink)  $YY$ interaction; 
$U^{N}_\Xi=-18$ MeV  (squares)  without (red) and with (blue)  $YY$ interaction: gravitational mass, baryonic mass, star radius, strangeness content and central density.}
\label{MR1}
\end{figure*}

The information related to the  maximum mass configurations presented in
Table~\ref{tab1} is partially plotted in Fig.~\ref{MR1}  for the parametrization TM1. We have considered
four values of $\Lambda_\omega$ corresponding to $L=55, 70, 80$ and 110 MeV in
order to be able to discuss the effect of the symmetry energy slope.

Some general conclusions may be drawn with respect to the strangeness
content:
a) if the strange mesons are not included, we get the smallest
masses 
 [see Table~\ref{tab1} and red ($U^N_\Xi=-18$ MeV), green ($U^N_\Xi=+18$ MeV) symbols in Fig.~\ref{MR1}a)]. A faster increase of the
strangeness content seems to be the reason for this behavior. These are also
the stars with the smallest strangeness content  [see  Fig.~\ref{MR1}d)];
b) the maximum star  mass changes with $L$, and stars with an intermediate $L$ have
the smallest masses and, generally, have the largest central densities [see
  Fig.~\ref{MR1}a) and e)].  There
are two competing factors that define this behavior: on one hand a larger $L$
corresponds to a harder EOS because the symmetry energy increases faster with
the density, on the other hand a larger $L$ favors larger strangeness
fractions which softens the EOS. The first one gives rise to smaller central
densities and larger radii, while the second one leads to the opposite; 
c) the strangeness content depends on the hyperon interaction, and, 
in particular, on the $\Xi$ potential in the present study. If $U^N_\Xi=+18$ MeV (triangles)
the masses are larger and the strangeness fractions generally smaller;
d) including the strange mesons gives rise to more massive stars which may have larger
strangeness contents. In this case the strangeness content is always smaller
for a smaller slope $L$, and its maximum value is of the order 0.04-0.05 according to the hyperon interaction if $L=55$ MeV. The upper limit can go to 0.07-0.08 if $L=110$ MeV,
Fig.~\ref{MR1}d). Larger fractions may be obtained if the $U^N_\Sigma$ is
considered attractive.

As discussed before, comparing stars with the same mass the
strangeness content depends on the hyperon-meson interaction, and, for instance, a 1.67 $M_\odot$
may have a strangeness fraction that goes from 0.005 to 0.115. The largest strangeness fraction was
obtained with $U^N_\Sigma=-30$ MeV, no strange mesons and $L=55$ MeV. For $L=110$ MeV there is no
stable star with this mass for these couplings.

We conclude that both
the symmetry energy and the strangeness content may give rise to similar
effects on some properties of the stars such as the radius.
 These two effects
  may be partially disentangled by analyzing stars in different mass ranges, 
since  hyperons only exist in  massive enough stars, in the
  present case only if $M> 1.5 \, M_\odot$. On the other hand low mass stars
  do not test the EOS at large densities. For TM1 the central density of a
  star with $M=1.2\, M_\odot$ is of the order of $2 \rho_0$.
The
effect on the radius of a  star with a mass $M=1.67\, M_\odot$ is similar  if  the slope $L$ decreases
from 110 MeV to 55 MeV or the 
 total strangeness fraction in the star increases from zero to 10\%.  In both
 cases the star radius suffers a decrease of 1 km.

In \cite{rafael11,providencia12} it was shown that a  smaller $L$
makes the nucleonic EOS softer giving rise to less massive stars with smaller
radius that varies approximately
linearly with $L$.  The inclusion of hyperons softens the
EOS but if the strange  mesons are included although the
strangeness fraction as a function of density  is smaller,
the total strangeness content in the star will be larger than when only non-strange
mesons are included. If hyperonic degrees of freedom are included the
radius of the star  decreases. An extra decrease of the radius occurs
when a smaller $L$ is considered, however, this  effect is
not so strong as in nucleonic stars.

An attractive or repulsive optical potential of the  hyperons, in the present study  $U^{N}_\Xi$,
clearly affects the fraction of strangeness as a function of density. A large fraction of
 strangeness at smaller densities makes the EOS too soft and smaller maximum masses and central densities
are obtained. This effect almost disappears if strange mesons are included.

At a given density the fraction of strangeness is smaller  for a softer
symmetry energy. 
 As a result, although the initial nucleonic EOS is softer, 
 a smaller  hyperon fraction  may  give rise to a slightly larger  maximum
mass,
 \textit{e.g.} 1.76 $M_\odot$ for L=55 MeV instead of 1.75 $M_\odot$ for L=110 MeV 
if $U^{N}_\Xi=18$ MeV or 1.69 $M_\odot$ for L=55 MeV   instead of 1.68 $M_\odot$  for L=110 MeV 
if $U^{N}_\Xi=-18$ MeV. This effect was also observed in the calculation 
with the QMC model \cite{panda12}. This is not anymore true when the strange mesons are included 
because it brings extra repulsion to the EOS.  
Therefore, an EOS with a larger strangeness content may become harder.

\begin{table*}
\caption{Properties of  maximum mass stars obtained with the TM1 \cite{tm1},
  TM1-2 and FSU \cite{fsu} nucleonic EOS and with the corresponding
hyperonic EOS with  $U^{N}_\Lambda=-28$ MeV and $U^{N}_\Sigma=30$ MeV and 
different values of $U^{N}_\Xi$.}
\label{tab1}
\begin{ruledtabular}
\begin{tabular}{ l l c c c c c c}
                  & $\Lambda_{\omega}$  & $M_{G}[M_{\odot}]$ & $M_{b}[M_{\odot}]$ & $R[\hbox{km}]$ & $E_c(\hbox{fm}^{-4})$ & $\displaystyle\frac{N_s}{\rho_B}$ & $\rho_c(\hbox{fm}^{-3})$ \\
\hline
no hyperons, TM1    & 0.0 & 2.18 & 2.54 & 12.39 & 5.36 & - & 0.85 \\
                         & 0.03 & 2.13 & 2.51 & 11.92 & 5.61 & - & 0.90 \\
no hyperons, TM1-2  & 0.0 & 2.29 & 2.69 & 12.57 & 5.15 & - & 0.81 \\
                         & 0.03 & 2.25 & 2.66 & 12.14 & 5.37 & - &
                         0.86 \\     
no hyperons, FSU        & 0.03 & 1.72 & 1.97 & 10.85 & 7.07 & - & 1.15 \\
                \\
 $U^{N}_\Xi=-18$ MeV \\                                                                                                                               
Without $(\sigma^*,\; \phi)$, TM1 & 0.0  & 1.68 & 1.88 & 13.51 & 3.81 & 0.0386 & 0.68 \\
                                    & 0.03 & 1.69 & 1.91 & 12.64 & 4.33 & 0.0395 & 0.77 \\ 
Without $(\sigma^*,\; \phi)$, TM1-2 & 0.0  & 1.74 & 1.95 & 13.52 & 3.86 & 0.0479 & 0.68 \\                                
                                    & 0.03 & 1.77 & 2.01 & 12.70 & 4.34 & 0.0486 & 0.76 \\                               
Without $\sigma^*$, TM1           & 0.0  & 1.88 & 2.13 & 12.40 & 5.34 & 0.0678 & 0.89 \\
                                    & 0.03 & 1.87 & 2.15 & 11.95 & 5.51 & 0.0535 & 0.92 \\                                                           
Without $\sigma^*$, TM1-2           & 0.0  & 1.96 & 2.23 & 12.50 & 5.26 & 0.0773 & 0.87 \\
                                    & 0.03 & 1.96 & 2.27 & 12.10 & 5.37 & 0.0617 & 0.89 \\                                                            
With $(\sigma^*,\; \phi)$, weak YY, TM1 & 0.0  & 1.77 & 2.00 & 12.72 & 4.87 & 0.0699 & 0.83 \\
                                          & 0.03 & 1.78 & 2.03 & 12.12 & 5.19 & 0.0594 & 0.89 \\
With $(\sigma^*,\; \phi)$, weak YY, TM1-2 & 0.0  & 1.84 & 2.08 & 12.76 & 4.90 & 0.0818 & 0.83 \\
                                          & 0.03 & 1.86 & 2.13 & 12.21 & 5.14 & 0.0694 & 0.87 \\
                                          \\
$U^{N}_\Xi=0.0$ MeV  \\                                                                                                                                                                                                                             Without $(\sigma^*,\; \phi)$, TM1 & 0.0  & 1.72 & 1.93 & 13.38 & 4.02 & 0.0412 & 0.71 \\
                                    & 0.03 & 1.74 & 1.97 & 12.62 & 4.39 & 0.0350 & 0.77 \\ 
Without $(\sigma^*,\; \phi)$, TM1-2 & 0.0  & 1.79 & 2.01 & 13.44 & 4.00 & 0.0484 & 0.70 \\                                
                                    & 0.03 & 1.82 & 2.07 & 12.75 & 4.33 & 0.0417 & 0.76 \\                               
Without $\sigma^*$, TM1             & 0.0  & 1.89 & 2.15 & 12.42 & 5.33 & 0.0628 & 0.88 \\
                                    & 0.03 & 1.89 & 2.17 & 12.00 & 5.44 & 0.0460 & 0.91 \\                                                           
Without $\sigma^*$, TM1-2           & 0.0  & 1.98 & 2.26 & 12.53 & 5.23 & 0.0712 & 0.86 \\
                                    & 0.03 & 1.98 & 2.29 & 12.16 & 5.28 & 0.0526 & 0.88 \\                                                            
With $(\sigma^*,\; \phi)$, weak YY, TM1 & 0.0  & 1.81 & 2.04 & 12.73 & 4.88 & 0.0636 & 0.83 \\
                                        & 0.03 & 1.81 & 2.07 & 12.19 & 5.10 & 0.0488 & 0.87 \\
With $(\sigma^*,\; \phi)$, weak YY, TM1-2 & 0.0  & 1.88 & 2.13 & 12.81 & 4.84 & 0.0730 & 0.82 \\
                                          & 0.03 & 1.90 & 2.18 & 12.33 & 5.00 & 0.0568 & 0.85 \\
                                          \\
 $U^{N}_\Xi=18$ MeV \\                                                                                                                                                                                   
Without $(\sigma^*,\; \phi)$, TM1 & 0.0  & 1.75 & 1.97 & 13.22 & 4.24 & 0.0468 & 0.74 \\
                                    & 0.03 & 1.76 & 2.00 & 12.51 & 4.58 & 0.0374 & 0.80 \\ 
Without $(\sigma^*,\; \phi)$, TM1-2 & 0.0  & 1.82 & 2.05 & 13.28 & 4.23 & 0.0542 & 0.74 \\                                
                                    & 0.03 & 1.84 & 2.11 & 12.65 & 4.51 & 0.0433 & 0.78 \\                               
Without $\sigma^*$, TM1           & 0.0  & 1.90 & 2.16 & 12.39 & 5.36 & 0.0622 & 0.89 \\
                                    & 0.03 & 1.89 & 2.17 & 11.99 & 5.44 & 0.0456 & 0.91 \\                                                           
Without $\sigma^*$, TM1-2           & 0.0  & 1.98 & 2.27 & 12.50 & 5.26 & 0.0701 & 0.87 \\
                                    & 0.03 & 1.98 & 2.29 & 12.15 & 5.29 & 0.0519 & 0.88 \\                                                        
Without $\sigma^*$, FSU             & 0.03 & 1.53 & 1.71 & 10.94 & 6.87 & 0.0348 & 1.16 \\

With $(\sigma^*,\; \phi)$, weak YY, TM1 & 0.0  & 1.83 & 2.07 & 12.63 & 5.03 & 0.0640 & 0.85 \\
                                          & 0.03 & 1.83 & 2.09 & 12.12 & 5.21 & 0.0479 & 0.89 \\
With $(\sigma^*,\; \phi)$, weak YY, TM1-2 & 0.0  & 1.91 & 2.17 & 12.71 & 4.97 & 0.0728 & 0.83 \\
                                          & 0.03 & 1.91 & 2.20 & 12.27 & 5.09 & 0.0548 & 0.86 \\
\end{tabular}
\end{ruledtabular}
\end{table*}

In this schematic study the largest mass obtained for hyperonic stars is 1.96 $M_\odot$,
compatible with  the mass of the pulsar  PSR J1614-2230, 1.97 $\pm 0.04$
M$_\odot$. However, there are still many uncertainties on the hyperon
interaction in nuclear matter and the EOS of nucleonic matter at
supra-saturation densities. It is, therefore,  not possible to take firm
conclusions with respect to the possible existence of hyperons  inside
compact stars, except that it seems important to include extra repulsion
between hyperons at high densities through the inclusion of strange mesons, as
was proposed in \cite{schaffner12a}.

\section{Conclusion}
\label{conclusion}
The density dependence of the EOS and the hyperon couplings have both a 
strong effect on the mass and radius of the star. 
We have tested  different hyperon-meson parametrizations, using
information from hyper-nuclei to fix the couplings, and different nucleonic
properties at supra-saturation density within the limits of experimental
constraints, namely the density dependence of the symmetry energy and the
incompressibility at large densities.

The chosen nucleonic EOS cover different density dependences of both the
isoscalar and isovector channels of the EOS. We have used an EOS with a
quite high incompressibility at saturation, $K=281$ MeV. However, the
non-linear vector meson quartic term  softens the EOS at
supra-saturation densities in such a way that the results from
Dirac-Brueckner-Hartree-Fock calculations are reproduced \cite{tm1}, and  the constraints
imposed by heavy ion collisions \cite{danielewicz} are satisfied. Also the slope of the symmetry energy
 of this model at saturation is probably too high according
to present experimental information. Therefore, we have tested other
EOS, with a softer symmetry energy and  with a harder EOS
at supra-saturation densities but still within the constraints of
\cite{danielewicz}. A softer EOS at saturation or above saturation
will not favor so much the onset of hyperons, therefore, smaller hyperon contents and larger hyperon onset
densities are expected. However, we should point out that a softer
EOS at saturation or just above will allow that larger central
densities are attained in the star, and, therefore, it is expected
that the exotic degrees like hyperons will be present in less massive
stars.

The present study estimates an upper limit of the expected hyperon content 
within RMF models taking into account the existing experimental constraints
 from heavy-ion collisions \cite{danielewicz}.

The effect of the density dependence of the  symmetry energy on low
mass neutron stars, $M<1.4 M_\odot$, with no hyperon content, has been 
discussed within the RMF formalism in \cite{hor03}. In the present study, 
we focus on hyperonic stars.
For the nucleonic EOS used, the hyperon degrees of freedom are present only 
for densities above 2.2 to 2.5 $\rho_0$, and inside stars with a mass larger 
than 1.5 $M_\odot$. However, using a softer EOS like FSU \cite{fsu},
hyperonic stars have a mass above 1.2 $M_\odot$. Since the EOS is softer,
matter is more easily compressed and exotic degrees of freedom occur in
less massive stars. In this case, the strangeness content in the most favorable
conditions discussed in the present paper does not exceed 0.035 of the total baryonic number in a
maximum mass configuration with $M=1.52\, M_\odot$.

We conclude that a softer symmetry energy gives rise to smaller stars with 
smaller strangeness content. The maximum gravitational and baryonic mass and
the central densities depend on the slope $L$ on a non linear way, and
intermediate $L$ values may give smaller masses and larger central densities.
It was also shown that for a star with a fixed mass the radius of the star
decreases linearly with the increase of the total strangeness content.
In particular a 1km decrease of the radius of a 1.67$M_\odot$ star may be
explained if the slope of the symmetry energy decreases from 110 to 55 MeV or the strangeness to
baryon fraction increases from zero to $\sim$ 0.09.

It was also shown that a softer symmetry energy corresponds to a slower increase 
of the hyperon fraction with density \cite{rafael11,panda12}. However, the onset 
of strangeness depends on the charge of the hyperons. Negatively charged hyperons 
set on at smaller densities while neutral hyperons appear at larger densities for
smaller values of $L$. If it is confirmed that the potential of
$\Sigma^{-}$ is repulsive in symmetric nuclear matter at saturation
we may expect that $\Lambda$ is the hyperon that arises at lower
densities, and, in this case, the onset of strangeness occurs at larger
densities with smaller slope $L$. 

For a nucleonic EOS we would expect that 
 the softer the EOS is, the larger are the central densities attained
and  the smaller the radius.  However,
hyperons will affect this simple relation and we have obtained larger
central densities with harder EOS when no exchange of strange mesons
or just a weakly attractive hyperon-hyperon interaction are considered. 
This is due to a slower increase of the hyperon content with density.

If the  $\sigma^*$ meson is not included and a repulsive hyperon-hyperon 
interaction is considered although a larger $L$ gives rise to a larger strangeness 
content, the extra repulsion between hyperons due to the presence of the $\phi$-meson 
compensates for the extra hyperon fraction and the effect of the symmetry energy is almost
not seen on the central density of the maximum mass configuration.
Using a harder EOS such as TM1-2 a larger hyperon fraction is
obtained for a given density. Due to the extra hardness and the
repulsive effect of the $\phi$ meson, matter is less compressible and stars 
with larger strangeness content are more massive.

The identified uncertainties will certainly affect the appearance of other
degrees of freedom, namely quark degrees of freedom, and we should expect a
transition to a deconfined phase at lower densities for a harder hadronic
EOS. The effect of the hyperon couplings and density dependence of the
EOS on the metastability of hyperonic matter to the conversion to
quark stars \cite{nucleation} should be investigated.

There is still lacking a lot of information about the nucleonic EOS at
supra-saturation densities as well as on the hyperon interactions in nuclear matter 
that may allow that an unambiguous  answer is given to the question whether the mass of the pulsar 
J1614-2230 could rule out exotic degrees of freedom from the interior of compact stars. 
We also conclude that some star properties are affected in a similar way by
the density dependence of the symmetry energy and the hyperon content of the
star. To disentangle these two effects it is essential to have a good
knowledge of the EOS at supra-saturation densities. Low mass stars will probably involve only nucleonic degrees of
  freedom and will allow to study  the density dependence of
  the symmetry energy effect alone. However, they will also only test
  densities not larger than
  2-3$\rho_0$.

\section*{Acknowledgments}
We would like to thank Jo\~ao da Provid\^encia for the reading of the manuscript and for many helpful and elucidating discussions.
This work was partially supported by COMPETE/FEDER and FCT (Portugal) under the grant
PTDC/FIS/113292/2009, and by COMPSTAR, an ESF Research Networking Programme.

\end{document}